\documentclass[prl,twocolumn,showpacs]{revtex4}

\def\gappeq{\mathrel{\rlap {\raise.5ex\hbox{$>$}}
{\lower.5ex\hbox{$\sim$}}}}

\def\lappeq{\mathrel{\rlap{\raise.5ex\hbox{$<$}}
{\lower.5ex\hbox{$\sim$}}}}

\newcommand{\lsim}{\raise.3ex\hbox{$<$\kern-.75em\lower1ex\hbox{$\sim$}}}
\newcommand{\gsim}{\raise.3ex\hbox{$>$\kern-.75em\lower1ex\hbox{$\sim$}}}
\begin{document}
\mathindent=0pt%

\title{Linking the parameters of diquark-quark model 
        to the Cabibbo angle  }%
\author{A.M.~Gavrilik}
\email{omgavr@bitp.kiev.ua}
\affiliation{Bogolyubov Institute for Theoretical Physics,
          03143 Kiev, Ukraine}
\author{I.I.~Kachurik}
\affiliation{Bogolyubov Institute for Theoretical Physics,
          03143 Kiev, Ukraine}

\date{\today}

\begin{abstract}
From two different modifications of the Gell-Mann-Okubo mass
relation for ${\frac12}^+$ baryons: the first one given
by a version of diquark-quark model and the second one being
an optimal mass sum rule obtained by using quantum groups
$U_q(su_n)$ for the role of hadronic flavor symmetries, we find
a direct connection of the mass parameters of the diquark-quark
model of Lichtenberg, Tassie and Keleman to the (proper
value of) $q$-parameter and then to the Cabibbo angle.
\end{abstract}


\maketitle


\def\a{\alpha }  \def\b{\beta }
\def\g{\gamma }  \def\G{\Gamma }
\def\d{\delta }   \def\D{\Delta }   \def\e{\epsilon }
\def\l{\lambda }  \def\L{\Lambda }  \def\k{\kappa }
\def\o{\omega }   \def\O{\Omega }
\def\t{\theta }   \def\S{\Sigma }
\def\ts{\textstyle}
\def\scr{\scriptsize}

           \section{Introduction}

One of the earliest improved or generalized versions of the
Gell-Mann - Okubo (GMO) mass relation for baryons forming the
$SU(3)$ octet  has been obtained by
Lichtenberg, Tassie and Keleman (LTK)
in a particular version of 'diquark-quark' model               \cite{LTK}.
The essence of the LTK result is that the correction to
GMO combination is expressed in terms of basic parameters
(of dimension of mass) characterizing both the diquark and the
third quark.
It is important to emphasize that viewing baryons as
consisting of a diquark and a separate third quark
provides a number of advantages                                \cite{APEFL}
among which one also finds the important ability
to account for some nonperturbative aspects of QCD.

On the other hand, quantum groups and quantum
($q$-deformed) algebras                                      \cite{Dr-Ji}               
provide a very useful tools not only for application
to the spectroscopy of diatomic molecules
and (super)deformed nuclei (see e.g.                           \cite{Chang})
but, as it was demonstrated more recently,
these are very useful when applied to phenomenological
description of hadron properties                               \cite{G_NPB}.
In the framework of the approach initiated in                  \cite{G-JPA}
(where the case of vector mesons was first considered)
and developed in more detail in subsequent papers, see          \cite{G_NPB}
and references given therein, the $q$-algebras 
$U_q(su_n)$ have been adopted, in place of the Lie
algebras of the groups $SU_n$, as those describing
flavor symmetries of hadrons. Such replacement allows
to derive numerous results concerning hadron masses
and mass sum rules, along with a number of interesting
implications.
Basic tool of this aproach is the representation theory
of the $q$-algebras $U_q(su_n)$                                \cite{Dr-Ji}.

In the case of baryons, it was clearly  
demonstrated that the approach takes into account, in
a uniform and natural  way, the contributions in baryon
masses which reflect essentially non-polynomial                 \cite{GI98}
(in fact, {\em all-order}) effects of $SU(3)$ breaking.
As another important consequence, the (phenomenologically)
most adequate fixed value of the deformation parameter $q$
is linked directly 
to the famous Cabibbo angle $\theta_{\scr{C}}$                 \cite{cabi}.

The goal of this note is to find a direct connection 
between the basic parameters involved in
the aforementioned two different extensions (modifications)
of the {octet baryon} GMO mass formula derived within 
the {\em apparently differing} approaches: from the evaluation 
of hadron masses using $q$-deformed counterpart  
of flavor symmetries $SU(N)$,
and from the diquark-quark treatment of baryons
within the LTK model.                                          

\section{ Baryon mass relation from the LTK 
          diquark-quark model } 

We first give a short sketch of those results from             \cite{LTK}
which are relevant for our further discussion.
With the notation
\[
\zeta = (m_s-m_t+v_s-v_t)/2V_0 ,
\]
\[
\g_t = ({\d_t -\d_q})/{6V_0} ,
\hspace{8mm}
\g_s = ({\d_s -\d_q})/{6V_0},
\]
(here the subscript $"s"$ or $"t"$ refers to $SU(3)$ sextet or
$SU(3)$ triplet diquark respectively, and $"q"$ refers to
the third quark), the octet baryon mass differences obtained in 
the LTK model are of the form                                   
\[        \hspace{-16mm}
m_{\L} - m_N =
{\ts \frac16 } (\d_t + 3 \d_s + 2 \d_q) - v_8
             \hspace{14mm}
\]
\begin{equation}
\hspace{16mm}
+ V_0\Bigl ( 2 \zeta (3 \g_s - \g_t)
+ 9 \g_s^2 - 6 \g_s\g_t + 5 \g_t^2  \Bigr ), \
\end{equation}
\begin{equation}
m_{\S} - m_\L =
{\ts \frac13 } (\d_t -\d_s ) - 4 V_0
\Bigl (
\zeta (\g_s + \g_t) + \g_s^2 - \g_t^2  \Bigr ), \
\end{equation}
\begin{equation}
m_{\Xi} - m_\S =                                   \label{differ}
{\ts \frac13} (2 \d_s + \d_q) - v_8 + 8 V_0\left (
\zeta \g_s + 3 \g_s^2 - 3 \g_s\g_t \right ). \  
\end{equation}
The parameters $\delta_t$,  $\delta_s$ and $\delta_q$
(of dimension of mass) in these expressions
are a measure of the violation of $SU(3)$ and, added properly to
the respective $SU(3)$ invariant masses $m_t$,  $m_s$ and $m_q$,
provide necessary mass splittings in the diquark triplet,
diquark sextet, and in the third quark $SU(3)$ triplet.
For further details concerning definition and physical meaning 
of all the involved parameters (including $v_s,\ v_t,\ v_8$
and $ V_0$) see ref.                                            \cite{LTK}.

The modified (improved) version of GMO relation obtained
in the LTK diquark-quark model is
\begin{equation}                                \label{C-LTK}
{\ts \frac32} m_\L
+ {\ts \frac12} m_\S - m_N - m_\Xi
= {\cal C}_{\rm LTK}    \equiv \mu_s \bigl (
3 \xi_{ts}^2 + 18 \xi_{ts} - 13 \bigr )     \hspace{0.6mm}
\end{equation}
where for convenience we have set  
\begin{equation}
\mu_s = V_0  \g_s^2 ,      \hspace{18mm}
\xi_{ts}\equiv \g_t /\g_s .    
\end{equation}
The quantity $\mu_s$ must be positive since
it can also be inferred by using
the decuplet mass combination                                  \cite{LTK}:
$8 \mu_s = 2 m_{\Xi^*} - m_{\S^*} - m_\O  = + 9.8\ $ MeV.
From the viewpoint of agreement with data, 
it is clear that there exists a continuum of values
for the pair $(\mu_s, \xi_{ts})$, determined in
the $\mu_s - \xi_{ts}$ plane by the curve
$\mu_s \bigl ( 3 \xi_{ts}^2 + 18 \xi_{ts} - 13 \bigr ) = {\rm const}$,
which provide the agreement of eq.(4) with data.
To reduce maximally such sort of non-uniqueness,
one needs some additional criteria.
To this end, LTK exploited the mass relation for
decuplet baryons.  
    Namely, taking the ratio of decuplet mass
combination to the octet one, then maximizing its r.h.s.
as a function of $\xi_{ts}$, they inferred the value
$\xi_{ts}=-3$                  
(remark, it is not clear why one has to just maximize).

Negative sign of the solution  $\xi_{ts}=-3$  implies that
the mass difference $\d_t$ must be greater (less) than $\d_q$ when
the mass difference $\d_s$ is less (greater) than $\d_q$.
However, this value $\xi_{ts}=-3$ is in conflict with
empirics:
it supplies negative value to the r.h.s. of      (\ref{C-LTK})
thus providing the correction to GMO
{\em in wrong direction}.

\section{ Quantum-group based baryon mass relations }

Now let us turn to another modification of the GMO 
mass sum rule, namely the $q$-deformed mass relation 
obtained, using the quantum algebras $U_q(su_n)$ taken 
for flavor symmetries, in the form                  \cite{Ga-Uzh,Breg,G_NPB}
\[
\vspace{-0.6mm}
[2]m_N+\frac{[2] m_{\Xi}}{[2]-1} = [3] m_{\Lambda }
+ \Bigl ( \frac{[2]^2}{[2]-1}-[3] \Bigr ) m_{\Sigma }                           
\]
\begin{equation}
\hspace{18mm}
+\frac{A_q}{B_q}\left( m_{\Xi } + [2] m_N -
   [2]m_{\Sigma } - m_{\Lambda } \right)       \label{q-MSR}
\end{equation}
with $[3]\equiv[3]_q=([2]_q)^2-1$, \
and with $A_q$, $B_q$ being certain polynomials in
$[2]\equiv [2]_q = q+q^{-1}$                         
whose sets of zeros are completely different.
This $q$-analog yields, as particular cases,
the familiar Gell-Mann - Okubo relation                          \cite{GMO}
$m_N+m_\Xi=\frac32 m_\Lambda+
\frac12m_\Sigma $
(known to hold with the $0.58\%$ accuracy)
at the 'classical' value   $q=1$, and
the whole infinite set of new mass sum rules        \cite{GI98,Ga-Uzh}
\[
m_N+{1\over [2]_{q_n}-1}m_{\Xi }=
{[3]_{q_n}\over [2]_{q_n}}m_{\Lambda }
+\Bigl ({[2]_{q_n}\over [2]_{q_n}-1}-
{[3]_{q_n}\over [2]_{q_n}}\Bigr )m_{\Sigma },
\]
\begin{equation}                                   \label{series}
          \hspace{2mm}           6\le n < \infty ,
\end{equation}
where $q_n = \exp (i\pi /n)$.
It should be mentioned that for each such value $q_n$ 
the respective sum rule shows better agreement with data 
than GMO one.

The phenomenologically most adequate mass relation 
among those contained in the series    (\ref{series}),
namely
\begin{equation}
m_N + m_{\Xi}/({[2]_{q_7}\!-\!1}) =                 \label{best}
      m_{\Lambda} /({[2]_{q_7}\!-\!1})+ m_{\Sigma} ,
\end{equation}
shows the remarkable $0.07\%$ accuracy.
This, most accurate, mass sum rule    
corresponds to the value
$q_7=e^{{\rm i}\pi/7}$, for which
a clear physical meaning was suggested  
in ref.                                                   \cite{Ga-Uzh,ARW}
where the value $q_7$ was directly linked
to the Cabibbo angle, i.e.
$\frac1i\ln{q_7} = \frac{\pi}{7}{=}2\theta_{\scr{C}}$.
   
Now let us present the optimal mass relation  (\ref{best})
in the form of GMO combination with a correction to it:
\begin{equation}                                  
   \hspace{-14mm}                            \label{C-q7}
{\ts \frac32} m_\L + {\ts \frac12} m_\S 
- m_N - m_\Xi = {\cal C}_{q_7} ,   
    \hspace{12mm}                     
\end{equation}
\[     \hspace{2mm}
{\cal C}_{q_7}
\equiv \left( 
             ({[2]_7-1})^{-1} - 1 \right)
(m_\Xi - m_\L) - {\ts\frac12} (m_\S - m_\L).     \ \ \
\]

Formulae            (\ref{q-MSR})-
                                         (\ref{best})
encode {\it highly nonlinear dependence} of mass on                                       
$SU(3)$-breaking.                        
This makes them  radically different from the classical GMO
result accounting only first order effects in $SU(3)$-breaking.

Such {\em nonpolynomiality} in $SU(3)$-breaking effectively 
accounted by the quantum-group based model, for the case of 
octet baryon masses was demonstrated in                         \cite{GI98}.
For this goal, the explicit dependence on
hypercharge $Y$ and isospin $I$ of matrix elements
for isoplet masses wss analyzed.
Typical contribution to octet baryon mass contains
such expressions as, e.g., the terms 
$\left({[Y/2]_q[Y/2\!+\!1]_q-[I]_q[I\!+\!1]_q}\right)$
or 
$\left({[Y/2-1]_q[Y/2-2]_q-[I]_q[I+1]_q}\right)$,
with multipliers depending on the labels
$m_{\scr{15}},m_{\scr{55}}$ of a chosen
dynamical representation.
This shows nontrivial explicit dependence 
on hypercharge and the factor $[I]_q[I+1]_q$ 
($q$-deformed $SU(2)$ Casimir).
Since the $q$-bracket is $[n]_q=\frac{\sin(nh)}{\sin(h)}$
if $q\!=\!\exp({\rm i}h)$, baryon masses depend 
on $Y$ and $I$ (that is, on $SU(3)$-breaking effects) 
in highly nonlinear - {\it nonpolynomial} - fashion.
The ability to account highly nonlinear
$SU(3)$-breaking effects, due to the use
of the quantum counterpart $U_q(su_n)$ of
usual flavor symmetries, is similar 
to the  result                                               \cite{Lor-We}
that by exploiting {\em appropriate free}
$q$-deformed structure one is able to efficiently
describe the properties of (undeformed)
quantum-mechanical system with complicated interaction.

\section{ Diquark-quark parameters and $q$-parameter}                                     

Now let us relate the results  (\ref{C-LTK})  and  (\ref{C-q7})
of two different approaches.
To this end we form, taking the mass differences
$m_\Xi - m_\L$ (multipled with some $w$) and  $m_\S - m_\L$
from (1)-(3), the particular combination involved
in the r.h.s. ${\cal C}_{q_7} $ of
(\ref{C-q7}) and equate it to the  r.h.s. of (\ref{C-LTK}).
Then, imposing 
\begin{equation}
4 w\                                             \label{10}
\mu_s \ (\xi_{ts}^2 - 6 \xi_{ts} + 5 )
    = \mu_s \ ( 5 \xi_{ts}^2 + 18\xi_{ts} - 15 ),
\end{equation}
\[     
w \bigl [ {\ts \frac13} (\d_t+\d_s+\d_q) 
- v_8 + 2 (\g_s-\g_t)(m_s-m_t+v_s-v_t) \bigr ]
\]
\begin{equation}       
 + {\ts \frac16}                                  \label{11}
(\d_s-\d_t) + (\g_s+\g_t)(m_s-m_t+v_s-v_t) = 0,
\end{equation}
with some $w$, guarantees validity of eq.(\ref{C-LTK}) and
its correspondence with eq.(\ref{C-q7}).
The solution of (10), namely
\begin{equation}
w = -1 +                                           \label{12}
\frac{9 \xi_{ts}^2 - 6\xi_{ts} + 5}
     {4 (\xi_{ts}^2 - 6 \xi_{ts} + 5 )} ,
\end{equation}
when put in (11) gives a particular constraint on
the parameters of the LTK quark-diquark model.

As result, we are led to the explicit relation between 
the (value $q_7 = \exp (i\pi /7)$ of)
$q$-parameter in our $q$-GMO and the ratio
$\xi_{ts}=\gamma_t/\gamma_s$ of LTK quark-diquark model:
\begin{equation}                                   \label{13}
 [2]_7 - 1 = 
\frac{ 4 (\tilde\xi_{ts}^2 - 6 \tilde\xi_{ts} + 5) }
{ 9 \tilde\xi_{ts}^2 - 6 \tilde\xi_{ts} + 5 } .    
\end{equation}
Note that the tilda over $\delta_t$, $\delta_s$ and 
$\delta_q$ in the relation  (\ref{13})  is put in order 
to indicate that these values are now the 
{\em optimized ones corresponding to all-order account} of
$SU(3)$ symmetry breaking in octet baryon masses, 
as encoded in the l.h.s. of (\ref{13}).

Since $[2]_7=2\cos{\pi\over 7}\cong 
1.80194$,
the obtained relation yields as its solutions
the two values   
namely
$ \tilde\xi_{ts}^{(+)} = 0.741 $  
and 
$\tilde\xi_{ts}^{(-)} = - 6.705 .$
By the very derivation, both these values guarantee
the validity of sum rule (\ref{C-LTK}) to within  $0.07 \% .$
The both values differ substantially from that
adopted by LTK and {\em provide positive correction to} GMO,
see the comment in sec.2 about the negative value of $\xi_{ts}$.
Moreover, our positive value $\tilde\xi_{ts}^+$
reflects (even qualitatively) different,
from the case of $\tilde\xi_{ts}^-$, physical situation.
This value implies that the mass difference 
$\tilde\d_t$ is greater (lesser) than $\tilde\d_q$ 
at the same time as the mass difference $\tilde\d_s$
is greater (lesser) than $\tilde\d_q$, since
\[
\tilde\d_t -\tilde\d_q = 0.741 (\tilde\d_s -\tilde\d_q),
\]
that is
\[
\tilde\d_t -\tilde\d_s = 0.259 (\tilde\d_q -\tilde\d_s) .
\]
Thus, it should necessarily be either
\begin{equation}                            \label{14}
\tilde\d_q > \tilde\d_t > \tilde\d_s
           \hspace{12mm}
\hbox{ or}
           \hspace{12mm}
\tilde\d_s > \tilde\d_t > \tilde\d_q .
\end{equation}
The deduced inequalities, we hope, give more realistic 
picture of the hierarchy among the parameters involved 
in the LTK diquark-quark model.

\section{ Linking the LTK model parameters to 
          the Cabibbo angle }

Now let us discuss the already mentioned connection 
between the $q$-parameter  and the  Cabibbo angle.
As it was shown in                                         \cite{Isa-Po},
the weak mixing is properly modelled by
the $q$-deformation.
On the other hand, there is the important relation 
$ \theta_{\scr{W}}
  = 2 (\theta_{12} + \theta_{23} + \theta_{13}),  
$
found in                                                     \cite{palle},
which connects $\theta_{\scr{W}}$ with the Cabibbo angle
$\theta_{12}\equiv \theta_{\scr{C}}$ (and the
$\theta_{13}, \theta_{23}$ mixings with 3rd family,
which we neglect).
This relates the mixing in  bosonic (interaction) 
sector with that in  fermionic (matter) sector of 
the electroweak model.
Combined with (8) this implies: the Cabibbo angle 
can be linked 
to $q$-parameter of a quantum-group (or $q$-algebra)
based structure applied in the fermion sector.
One can thus infer                                        \cite{ARW,G_NPB}
\begin{equation}
\theta_{\bf 8} =
\frac{\pi}{7}=2~\theta_{\scr{C}} .        \label{(15)}
\end{equation}
The latter formula suggests for Cabibbo angle 
the exact value $\frac{\pi}{14}$.
(Remark that for the $q$-deformed analog of decuplet 
mass formula                                                \cite{GKT,ARW},
we have $\theta_{\bf 10}=\theta_{\scr{C}}$.)

As a final result of this note we deduce the following          
corollary concerning 
%
{\em direct link of the Cabibbo angle to the (optimized) 
parameters of the diquark-quark model.}
%

Indeed, using 
$\theta_7 = \frac1i\ln{q_7} = 2 \theta_{\scr{C}}$ 
as given in (15), we obtain from eq.(13) the formula 
in question, that is
 \begin{equation}                               \label{final}
\cos{2 \theta_{\scr{C}}}
= \frac{13\tilde\xi^2_{ts}-30\tilde\xi_{ts}+25}
         {2(9{\tilde\xi}^2_{ts}-6{\tilde\xi}_{ts}+5)},
   \hspace{8mm}
\tilde\xi_{ts}\equiv\frac{\tilde\delta_t-\tilde\delta_q}
                         {\tilde\delta_s-\tilde\delta_q} , \ \
\end{equation}
This remarkable formula connects the Cabibbo angle with
the parameters of the LTK model, characterizing the
diquark and the third quark, whose optimized values are 
such that the inequalities (14) should hold.

\section{ Appendix:  
          LTK model parameters and $q$-parameter,
          general case}    

In the appendix we demonstrate that, in principle,
the connection with the result (\ref{C-LTK}) of 
LTK model can be founded {\em in the general case} of 
an arbitrary member from the infinite discrete set 
presented in        (\ref{series}).

Indeed, eq.(\ref{series}) can be rewritten as
\[
{\ts \frac32} m_\L + {\ts \frac12} m_\S - m_N - m_\Xi =
\Bigl(\frac{[3]_{q_n}}{[2]_{q_n}}-\frac12 \Bigr ) (m_\S - m_\L)
\]
\vspace{-3mm}
$$
  \hspace{-8mm}
+  \frac{1}{[2]_{q_n}-1} (m_\Xi - m_\S)
- m_\Xi + m_\L.                                      \eqno(7')
$$
%
Now take mass differences from (1)-(3).
Multiply $m_\Xi - m_\S $ with some $x^{-1},\ $   $m_\S - m_\L $
with the appropriate \footnote{ Just this explicit dependence 
of $y$ on $x$ is motivated by the fact that, in the sequel, 
$x$ is to be juxtaposed with $[2]_q-1$, while the $q$-number 
$[3]_q$ is expressible in terms of $[2]_q$ as 
$[3]_q=([2]_q+1)([2]_q-1)$. }
\[
y = \frac{x(x+2)}{x+1}-\frac12,
\]
then form the particular combination that corresponds
to the r.h.s. of ($7'$) and equate it to the r.h.s.
of (4).
The outlined procedure gives us: \ the relation
\[
4 \Bigl (\frac{x(x+2)}{x+1}-\frac12 \Bigr )
 \left ( \xi_{ts}^2 -1 \right ) 
        - 24 x^{-1} (\xi_{ts} -1)
\]
\begin{equation}                               \label{17}
          \hspace{-26mm}
= 7\xi_{ts}^2 - 6 \xi_{ts} + 7
               \hspace{12mm}
\end{equation}
and the additional constraint
\[
\Bigl (\frac{x(x+2)}{x+1}-\frac12 \Bigr )
\Bigl \{\frac13 (\d_t-\d_s) 
- 4 \mu_s\zeta (1+\xi_{ts}) \Bigr \}
\]
\[                       \hspace{-12mm}
+ x^{-1} \Bigl \{\frac13 (2\d_s-\d_q) 
   - v_8 +8 \mu_s\zeta \Bigr \}   \hspace{6mm}
\]
\begin{equation}
           \hspace{6mm}                         \label{18}
- \frac13 (\d_t+\d_s+\d_q)
  + v_8 + 4 \mu_s\zeta (\xi_{ts}-1) = 0  \hspace{-2mm}
\end{equation}
with $\xi_{ts}$ given in (5),
when hold simultaneously, provide validity of eq.(4). 
Moreover, the latter is directly correspondent with 
the eq.($7'$) if the identification
\[
  x \ \longleftrightarrow \ [2]_{q_n}\!-\!1
\]
is made.
Not going into further details, let us only mention
that just at $q=\exp(i\pi /7)$ (i.e., in the case most adequate
phenomenologically) we have the equality
\[
[3]_{q_7}/[2]_{q_7}=([2]_{q_7}-1)^{-1},
\]
or, put in another way,
\[
\frac{x(x+2)}{x+1} = x^{-1}.
\]
With this simplification, we recover the relations (12)-(13) 
of the distinguished particular case considered above.
It is this remarkable phenomenological validity of (8) or (9) 
that makes it possible to link the LTK diquark-quark model 
parameters to the Cabibbo angle, as expressed in eq. (16).

\end{document}